# Designing a Network Based System for Delivery of Remote Mine Services

Craig James, Weidong Huang, Kazys Stepanas, Eleonora Widzyk-Capehart, Leila Alem, Chris Gunn, Matt Adcock, Kerstin Haustein
CSIRO, Australia
Email: craig.a.james@csiro.au, tony.huang@csiro.au, kazys.stepanas@csiro.au, eleonora.widzyk-capehart@csiro, leila.alem@csiro.au, chris.gunn@csiro.au, matt.adcock@csiro.au, kerstin.haustein@csiro.au

*Abstract*—There is a great body of work in the areas of tele-assistance/tele-collaboration offering novel and effective ways to improve collaboration between personnel located at a remote mine site and off-site personnel located in major metropolitan areas. Much of this work involves the use of high-bandwidth communications or targeted sensory experiences using large format displays. There are also existing remote access technologies but these suffer from limited functionality (providing text, voice, video or one-way desktop sharing), are often poorly supported in the security-conscious corporate environment and require complicated set up processes. There is currently no singular piece of remote collaboration technology that is suitable for the delivery of high-quality planning and scheduling services to clients at a mining site from a remote operating centre. In response to this issue, as part of a research and technology development effort between CSIRO and a mining engineering firm, we have developed a concept of remote mining engineer (RME) and conducted a functional requirements analysis for delivering mining engineering services to mine sites remotely. Based on the obtained requirements, a further study was performed to characterise existing technologies and to identify the scope for future work in designing and prototyping a network based system for RME. We report on the method and findings of this study in this paper.

## I. INTRODUCTION

Mining engineers play a critical role in on-site mine planning and operations but access to skilled staff willing to work in remote locations is difficult [8, 9]. While mining companies (service requesters) and service providers based in metropolitan centres are able to attract and retain top-level personnel, the need to ensure effective communications between off-site and on-site personnel, requires frequent trips to remote mining locations, which results in high travel burdens and high service costs. Further, after a site visit, the contracted mining engineer continues to engage with on-site personnel via emails or phone calls to ensure the services requested are delivered (such as short term mine planning). The time spent travelling to mine sites and collaborating through phone calls and emails is costly, time consuming and inefficient.

On the one hand, remote communication technologies such as tele-conferencing, Skype, desktop sharing [1] and even regular phone and email services can be utilised as a means of communication. However, individually, these technologies do not address the following key challenges in the open cut mining environment:

- Quality of service – effective remote communication relies on clear reception of as many cues as possible (text, tone, gesture, facial expressions) to avoid misunderstandings. Current applications of remote technologies do not provide a sufficient mix of these cues at a sufficient level of quality.

- Low bandwidth – broadband communications in remote areas is still very poor and the scale of information being shared (voice, text, data) has rapidly out-stripped available bandwidth

- High security – great care must be taken to protect integrity of data and control systems where downtime from malicious intrusions can introduce high production penalties

- Usability – available remote technologies are difficult to set up, configure and maintain, often requiring specialist resources and training.

On the other hand, there is a great body of work in the areas of tele-assistance/tele-collaboration that are specifically designed to improve collaboration between personnel located at a remote mine site and off-site personnel located in major metropolitan areas (e.g., [3]). Much of this work involves the use of high-bandwidth communications or targeted sensory experiences using large format displays [4].

Generally, there is currently no singular piece of remote collaboration technology that is suitable for the delivery of high-quality planning and scheduling services to clients at a mining site from a remote operating centre. In response to this issue, the concept of Remote Mining Engineer (RME) has been investigated through collaboration between CSIRO and a mining engineering firm (service provider) [2, 5]. The ultimate objectives of this collaboration project include:

- Facilitate collaboration between mining engineers inside the service provider and between staff of the service provider and the service requester. These participants are often physically distributed.

- Reduce the need for mining engineers to be present at mine sites without compromising the quality of mine planning and scheduling. This will be achieved by dramatically improving the communication ability between off-site and on-site personnel through technological innovations.

The customised RME system would combine existing (text, voice, visualisation and data sharing) and innovative communication technologies (tele-presence, tele-collaboration, tele-assistance, and immersive environments). These technologies will be combined in a way that improves collaboration and communication over long distances between on-site and off-site personnel, allowing service providers to deliver their mining engineering expertise to remote mine sites from their offices located in major metropolitan centres so that travel costs can be reduced without compromising the quality of the service. More specifically the system will rely on the following technologies:

- Tele-presence technology to enable a sense of physical presence of the remote mining engineer within mine site personnel.
- Collaborative workspace and desktop/whiteboard sharing to enable site personnel to: see what the mining engineer is working on, control the mouse of the remote PC, manipulate the CAD data in collaborative manner, and share ideas using whiteboard to write notes and produce sketches.
- Communication technologies (video and audio) to enable information exchange between the remote mine site and the service provider office.
- Visualisation technology to share two and three-dimensional mining data.

In the remainder of this paper, we briefly introduce the work we have done for requirement analysis. Then our methods and findings of reviewing existing technologies and recommendations for RME design and prototyping are presented. The paper concludes with a short summary and future work.

## II. BACKGROUND

### A. Model of Remote Service Delivery

Our previous studies revealed that service providers often follow a largely common business model to deliver remote mine services to their clients [2]. This model is illustrated in Figure 1.

As shown in Figure 1, a business case is started by a service request from the client. Upon the receipt of the request, a range of work routines follow on the service provider side. These include project initiation, site visit/s, internal task assignment, task progress reports and checkups, task collaboration and discussion, document exchange and task reassignment. Depending on the context of the request, the service can be executed by one or more mining engineer either from the same site or from different locations. The communication methods used by people involved include face-to-face, phone calls, one-line video, audio and text discussions, emails, sharing and sending data via physical media.

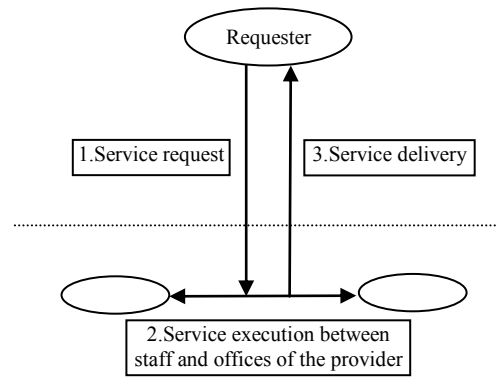

Figure 1. Business model.

### B. Requirement Analysis

Observation-based user experience design methods were combined with scenario-based software design techniques for requirement analysis. In particular, meetings were conducted with the service provider manager to understand the high-level expectations and identify typical work procedures and scenarios. On-site observations were undertaken of how work is actually carried out by engineers and, subsequently, focus-group meeting and individual interviews were carried out to elicit and analyse user needs.

### C. Key Requirements

The results of our requirement analysis highlighted three key challenges in developing remote collaboration software for the project, namely:

1. Bandwidth limitations.
2. Security concerns.
3. Usability issues.

Seven user cases were identified as key requirements when implementing a solution for delivering effective services remotely. These user cases are:

1. Ability to communicate through text, audio and video.
2. Ability to send a request to use the system and accept / reject it.
3. Ability to manage the system and configuration of users.
4. Ability to share full screens or application windows.
5. Ability to manage different parts of the collaboration tool over different screens.
6. Ability to share electronic whiteboards and annotate over applications.
7. Ability to transfer data over the network.

## III. METHOD

Based on the user requirements obtained, it was decided to make most use of existing technologies for design and development of the RME system. As a result, a scoping study was planned to identify the best suitable technologies to build a knowledge base for the project team and to inform the design of the RME system. It is important to note that our scoping study was not the functionally or useability test of those technologies or products.

*A. Identify Suitable Candidates*

There are a large number of tele-collaboration products on the market that offer a range of services that had the potential to service some of the requirements for delivering services to remote clients.

A broad product survey was conducted, identifying 56 possible candidates in the market. These 56 candidates were then further examined. This examination was expected to provide a recommendation on an initial system that would satisfy some of the requirements, and highlight scope for extensions or replacements in further implementation stages, in order to satisfy all of the requirements identified.

The scoping study narrowed the candidate list down to 6 likely products, which were then tested against some of the more prominent limitations and user cases required. Out of these tests, 2 candidates were highlighted as products that satisfied some of the limitations and user cases required.

Of these 2 products, Adobe's Connect Pro [6] and Scisco's WebEx [7] were highlighted. Connect Pro was felt to be more suitable as it was more stable and handled 3D content better. A detailed summary of this process is included in the next section.

*B. Presenting Results*

The results of the scoping study, a live demonstration of Connect Pro, a recommendation to use Connect Pro in the short term and design recommendations for the RME system were presented to the project stakeholders during a visit to the mining engineering firm.

This visit also included discussions on future work. It was agreed for CSIRO and the firm to purchase copies of Connect Pro, with the firm to try out using the product between their offices that were located in different cities, and CSIRO to incorporate the product in other strategic research to build up further knowledge.

## IV. RESULTS

*A. Broad Review*

The first step was to survey the available collaboration tools and decide on the best one that would immediately be usable for mining engineers to work together across a network. To do this, a list of 56 possible contenders for this project was compiled.

With the wide variety of products it was necessary to narrow the key types of collaboration that would be needed and the relative importance of each of these. These features, in order of importance were:

1. Audio communication. Without audio, nothing can happen. It was also assumed that the package itself would need to provide the audio connection, since a telephone line may be too expensive or inconvenient.
2. Sharing of snapshots of applications.
3. Annotation by all participants over those shared snapshots.
4. Sharing of live applications. It was assumed that these applications (such as a mining CAD package called Vulcan) would use OpenGL for 3D display. This is important because several products could share applications but not those using OpenGL.
5. Sharing control of live applications.
6. Video of participants.
7. Ability to transfer files.
8. Recording.
9. Having a tool to schedule meetings.

Many of these products were eliminated because they did not have video or audio communication, could not share applications, had poor security facilities, could not deal with network firewalls, or otherwise did not sufficiently satisfy enough of the limitations or user cases required by the project.

*B. Narrow Review*

This list of 56 names was then pared down to a shortlist of 8, based on the product descriptions and documentation. These products in the shortlist, including JoinMe, YuuGuu, MeetingPlace, GoToMeeting and Mikogo, TeamViewer, Connect Pro and WebEx, were then downloaded, installed and tested.

The first 5 were eliminated because they did not have appropriate audio and/or video facilities. The sixth, TeamViewer, was eliminated because it had problems with annotation over OpenGL applications (such as Vulcan). The main problem with sharing OpenGL applications was that the OpenGL refresh would immediately erase or damage any annotations that had been overlaid.

*C. Use-case Testing*

The remaining two, WebEx and Connect Pro were investigated in more detail. These two products were installed and tested with a full set of test cases developed based on the user requirements. It was found that these products were very close in the features that were offered. The main difference was in the way they responded when annotation modes were selected:

- Connect Pro took a snapshot of an application and participants drew over the top of this fixed image (see Figure 2).
- WebEx allowed participants to annotate in the same way over a window that could continue to change (such as a 3D animating application).

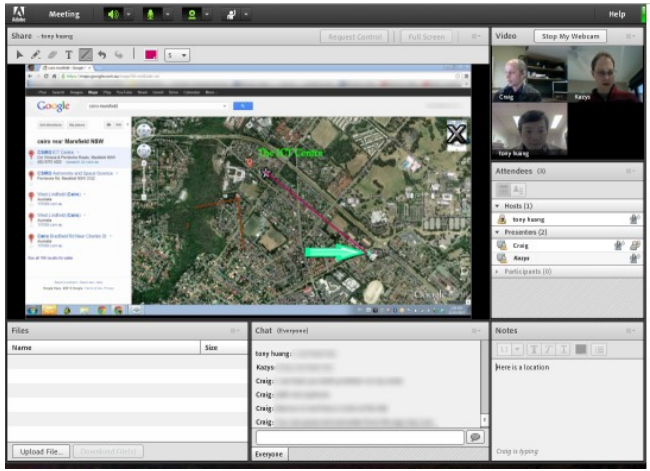

Figure 2. Adobe Connect used for sharing and group annotation.

The second approach seemed more powerful as it would allow for indicators, such as animating lines or pointers, to continue operating in a scene. However, the live annotation feature in WebEx can sometimes be unreliable with 3D applications, with annotations disappearing in many situations. As a whole, Connect Pro appeared to more preferable than WebEx in meeting our user needs.

*D. Bandwidth Testing*

Connect Pro was then tested against a key project limitation, reduced bandwidth conditions. For this test, a bandwidth-throttling program was installed at one end of a high-bandwidth network to simulate various low-bandwidth conditions.

It was assumed that reliable audio communication was essential for any collaboration, so this was taken as a measure of usability. If the audio became unusable, the condition failed. Tests were repeated for different combinations of features, to find the minimum bandwidth that could sustain the features and still permit understandable bi-directional audio. The results are shown in Table 1.

TABLE I. BANDWIDTH TEST RESULTS

| Bandwidth (Kbits/s) | Audio | Video | Sharing | Annotation | Experience[1] |
|---|---|---|---|---|---|
| 100 | Y | N | N | N | Poor |
| 150 | Y | N | N | N | Good |
| 150 | Y | N | Y (low)[2] | Y | Good |
| 150 | Y | N | N | Y | Good |
| 350 | Y | Y | N | N | Good |
| 350 | Y | Y | Y (low)[2] | Y | Ok |
| 400 | Y | Y | Y (low)[2] | Y | Ok |
| 450 | Y | N | Y (high)[3] | Y | Good |
| 525 | Y | Y | Y (high)[3] | Y | Good |

[1] – Subjective experience rating; poor, Ok, good.
[2] – Low refresh frequency OpenGL application (Qt sample).
[3] – High refresh frequency OpenGL application (Google Earth).

The shared application used for this test was Google Earth, as it uses OpenGL for its 3D rendering. During the tests, audio was considered unusable if parts of the audio stream were missing or if the latency was so large that conversations were not possible. It was observed that as the bandwidth was limited, the audio latency would increase. This may be due to packet retransmission within the TCP/IP communications mechanism. During two-hours of the bandwidth testing session the overall upload and download data transfer was greater than 1 Gigabyte each way. This may illustrate that the overall throughput allowance must be fairly high regardless of bandwidth.

It should be noted that Connect Pro did not automatically detect bandwidth and adjust any features. These had to be turned on and off by users.

*E. Results Summary*

Connect Pro was chosen for the more explicit style of sharing and annotating over an application. It was clearer that a snapshot view was being annotated upon, and it left the presenter with the ability to interact with other windows on the desktop when annotation is enabled. Also with Connect Pro, there were no irregularities with annotation on any of the windows tested.

V. DESIGN RECOMMENDATIONS

Generally speaking, existing technologies fall into three broad categories:

- Video conferencing
- Remote application and desktop control
- Web conferencing

Connect Pro was recommended as the top-matching product of a very wide field. Its strength highlighted in our study is the coverage it offers in all three of these categories. Although it is a powerful collaboration tool, Connect Pro has several severe shortcomings when contrasted against the project requirements. That said, Connect Pro provides a software development kit and many plugins have been developed that can extend and even replace existing functionality with targeted solutions. This opens up many opportunities for making improvements on current disadvantages.

*A. Bandwidth and Latency*

Bandwidth is generally viewed as an ever-increasing commodity (see Table 2). As such, the attention needed to address bandwidth concerns may often be overlooked. While bandwidth does continue to increase, demand is also set to increase.

TABLE II. PREDICTED IP TRAFFIC GROWTH 2014-2016 IN PETA BYTES. SOURCE: CISCO VISUAL NETWORKING INDEX: FORECAST AND METHODOLOGY 2011-2016 [10]

| IP Traffic, 2011-2016 | 2014 | 2015 | 2016 |
|---|---|---|---|
| By Geography (PB per Month) | | | |
| North America | 19,796 | 23,219 | 27,486 |
| Western Europe | 16,410 | 20,176 | 24,400 |
| Asia Pacific | 24,713 | 31,990 | 41,105 |
| Latin America | 3,495 | 5,208 | 7,591 |
| Central and Eastern Europe | 3,196 | 4,419 | 5,987 |
| Middle East and Africa | 1,417 | 2,320 | 3,714 |

In the case of the RME system, mine sites generally have poor internet connections due to security concerns and physical isolation. Even with high bandwidth infrastructure in place, other factors may reduce the available bandwidth and/or introduce latency. Network security may introduce latency, while high concurrent usage can reduce bandwidth – especially on WiFi networks – and mobile internet connections continue to lag behind bandwidth capabilities of fixed infrastructure. In these cases, bandwidth variability is the primary issue.

As such Quality of Service guarantees that cater for bandwidth and latency tolerance should be an essential part of any web-based communication product. At the very least, variable audio/video shaping technologies similar to those implemented by Skype should be included, catering for both low and variable bandwidth situations. Not all products reviewed provide this adaptability, choosing instead to measure available bandwidth only on initialisation.

*B. Security*

Connect Pro uses SSL encryption to secure communications. However, these connections are routed through central Adobe servers. Self-hosting options exist for enterprise level subscriptions at an additional cost. While this ensures that communications are adequately secured, some situations may require more direct connections, or the expense of self-hosting may not be justifiable.

Desktop sharing allows participants to see the host's desktop. Connect Pro supports restricting visibility of individual windows. We view this as an essential security features. However, some rendering artefacts are visible over hidden windows (garbled graphics).

*C. Programmatic Enhancements*

During testing it was found that while Connect Pro offered a wide variety of features, certain features were missing or needed improvement.

The most technical of these is the ability to annotate over live, 3D windows using a high refresh rate. Connect Pro's approach of capturing an image of the screen and annotating over this static image provides a very straightforward workaround, but was found to be inconsistent. Often different clients showed different frames captured from animated windows. In this context, annotations may not make sense to some participants, as they cannot see the intended target image. Annotation over live windows offers a more dynamic experience and allows the simultaneous integration of several features – such as annotation and remote application control.

Below is a list of other desirable features and their justifications.

- *Guaranteed connection*: Connect Pro attempts to maintain a connection in low-bandwidth environments but was still prone to dropping connections when communications was lost entirely. This required the user to reinitiate a connection and renegotiate a remote connection each time. After initiating a connection, links should continually attempt to re-establish broken connections until the user manually disconnects.

- *Virtual cursor*: each participant controls a virtual mouse pointer. This allows transient gestures and features highlighting without needing to manage annotations.

- *Control request*: where remote application control is allowed, participants may request control of an application using integrated features rather than vocalisation. This forms a more formal and unobtrusive request mechanic.

- *Improved file transfer*: file transfer should not be limited to the current session. That is, file transfers initiated during a session may complete after the session has been terminated. This allows larger file transfer. File transfer must also be integrated with bandwidth and latency tolerance algorithms.

- *Private whiteboards*: private chat features already exist, allowing users to send private text messages. This concept is extended to private whiteboards, allowing sharing of images, documents, drawings and annotations between subsets of participants.

- *User/client database and management*: at the time of evaluation, there was no address book built in to Connect Pro. User management and configuration tools allow for better management of repeated connections, meeting scheduling and user privileges. Better still, integration with an external client database maximises the management of clients in a single location.

- *Layout management*: Connect Pro features many internal windows including video, chat, user lists, annotation and presentation, etc. Layout management, restoration and sharing may enhance productivity and familiarity.

VI. CONCLUDING REMARKS AND FUTURE WORK

In this paper, we have presented our approach towards the design and implementation of the RME system. This approach makes use of the existing the technologies for knowledge development and for system design. First, 56 candidates were compiled and compared based on their functionalities and application requirements. This resulted in 8 products being identified for further testing in a simulated mining office environment. In the end, Connect Pro was the winner that was considered to be the best suitable system to meet our specific user needs and to inform the design and implementation of the RME system. During the process, additional functions were also proposed in addition to what has been available in Connect Pro.

For future work, we plan to start the full system development life cycle based on the obtained user requirements and design recommendations. It is hoped that end users will be fully involved in the process and their needs will be fully addressed whenever possible. We also plan to experiment and incorporate some additional technologies into the RME system. These include augmented reality, remote gestures, remote fault diagnosis and virtual presence.